\documentclass[prl,twocolumn,showpacs,floatfix]{revtex4}
\usepackage{amsmath}
\usepackage{amssymb}

\usepackage{graphicx}
\usepackage{dcolumn}
\usepackage{bm}

\begin{document}

\title{Role of interactions in the far-infrared spectrum of a lateral quantum
  dot molecule} 

\author{M. Marlo}
\author{A. Harju}
\author{R. M. Nieminen}
\affiliation{
Laboratory of Physics, Helsinki University of Technology, P. O. Box
1100 FIN-02015 HUT, Finland   
}

\date{\today}

\begin{abstract}

We study the effects of electron-electron correlations and confinement
potential on the far-infrared spectrum of a lateral two-electron
quantum dot molecule by exact diagonalization.
The calculated spectra directly reflect the lowered symmetry of the
external confinement potential.
Surprisingly, we find interactions to drive the spectrum towards that
of a high-symmetry parabolic quantum dot. 
We conclude that far-infrared spectroscopy is suitable for probing effective
confinement of the electrons in a quantum dot system, even if
interaction effects cannot be resolved in a direct fashion.  

\end{abstract}


\maketitle

\date{\today}

\maketitle

Nanoscale semiconductor structures are very promising for future
components of microelectronic devices.  These systems are also
scientifically very interesting as they exhibit novel and fundamental
quantum effects.
The most prominent difference between the two-dimensional artificial
atoms, or quantum dots (QD), and their normal counterparts are the
enhanced correlation and magnetic field effects.
In addition to that, the new features can be controlled by the tunable
system parameters, both experimentally and in theoretical models.
Coupling together QDs, one can construct QD molecules (QDM). The
electronic structure of these is also quite intriguing.
For example, the
two-electron QDM has a highly non-trivial spin-phase diagram and
composite-particle structure of the wave function~\cite{Ari}.
One experimental possibility to probe the new physics of
nanostructures is use far-infrared (FIR) absorption
spectroscopy~\cite{HeitmannKotthaus, JacakQD}.  For highly symmetric
QDs, the FIR spectra reflect only the center-of-mass properties of the
system. If the symmetry is lowered, the relative motion of the
electrons starts to couple to the center-of-mass motion, enabling one
to study the many-body correlation effects using FIR.

In this Letter we present FIR spectra for a two-electron QDM using
exact diagonalization.  The method used solves accurately the
quantum-mechanical model for a QDM, revealing the important
electron-electron correlation effects.
Our calculated spectra show clear deviations from the spectra of 
highly symmetric QDs. To analyze these deviations,
we vary our system parameters in order to separate the correlation
effects from the single-particle effects of the low-symmetry
confinement potential. This analysis shows, however, that these
effects are ultimately entangled. Furthermore, we find, very
surprisingly, that the correlation effects actually compensate the
lowered symmetry of the QD confinement potential, resulting in spectra
much closer to an electron in a high-symmetry QD than in a
low-symmetry one.

The starting point for understanding FIR spectra is the
highly symmetric, parabolic QD. This spectrum consists of two Kohn
modes, whose dispersion does not depend on the number of confined
electrons nor their interactions~\cite{MaksymChakra}.
Several experiments on QDs have shown deviations of the FIR spectra
from the Kohn
modes~\cite{HeitmannKotthaus,Demel,Dahl,Dahl2,Hochgrafe}.
It is clear that these deviations require a non-parabolic QD, but the
detailed cause of the deviations, and thus the interpretation of
the measured spectra, is difficult to obtain.
An especially interesting question is how the electron-electron
interaction and correlation effects emerge in the FIR spectrum when the
symmetry of the QD is lowered. 
In our model, we can tune both the deviation of the potential from being
perfectly parabolic as well as the electron-electron interaction
strength, which enables us to analyze the source of the non-trivial
details seen in the FIR spectra.  This type of analysis is extremely
important for explaining the details of the spectra.

The calculated FIR spectra have shown that a
broken $xy$-symmetry ({\it e.g.} elliptic or rectangular) is required
to observe the zero-field splitting of the Kohn modes as well as
additional modes, and a broken rotational symmetry ({\it e.g.} square)
is required to observe
anticrossings~\cite{Veikko,Pfannkuche,Madhav,Magnusdottir,Ullrich,Rodriguez}.
However, the role of interaction effects and lowered QD symmetry
in FIR spectra is not fully analyzed in these studies.
Our QDM model suits perfectly for this detailed analysis as the
FIR spectra show all the above-mentioned features.
One interesting finding is 
that the electron-electron interactions drive the FIR spectrum closer
to the Kohn modes instead of making the deviations more pronounced.
This means that the electrons feel an effective potential that is more
parabolic than the the bare confinement one. Thus the interactions
smoothen the non-parabolicity of the external potential.

We model the two-electron QDM by a 2D Hamiltonian
\begin{equation}
H = \sum _{i=1}^2\left ( \frac{ ( {- i {\hbar} \nabla_i}
-\frac ec \mathbf{A})^2 }{2 m^{*}} + V_\mathrm{c}({\bf
r}_{i}) \right ) + 
C \frac {e^{2}}{ \epsilon   r_{12} } \ ,
\label{ham}
\end{equation}
where $V_\mathrm{c}$ is the external confining potential, for which we
choose $\frac 12 m^* \omega_0^2 \min \{(x-d/2)^2+y^2,
(x+d/2)^2+y^2\}$.  This potential separates to two QDs at large
inter-dot distances $d$, and with $d=0$ it simplifies to one parabolic
QD.  We use the GaAs material parameters $m^*/m_e=0.067$ and
$\epsilon=12.4$, and the confinement strength $\hbar\omega_0=3.0$~meV.
$\mathbf{A}$ is the vector potential of the magnetic field (along the
$z$ axis) taken in the symmetric gauge. We use $C$ to scale the
Coulomb interaction strength.  The wave functions and energies of the
ground and excited states are found by the exact diagonalization of
the Hamiltonian matrix as in Ref.~[\onlinecite{Ari}].

Using the Fermi golden rule within the electric-dipole approximation
for the perturbing electromagnetic field, the transition probability
from the ground state to the $l$'th excited state can be calculated as
\begin{equation}
\mathcal{A}_{l, \pm} \propto  \left|  \left< \Psi_{l}
\left| e^{\pm  i \phi} \textrm{$\sum _{i=1}^{2}$}   
\mathbf{r}_i \right| \Psi_{0} \right> \right| ^2
\delta(E_l-E_0-\hbar \omega) \ ,
\end{equation}
where we have used right- or left-handed circular polarization of the
field, corresponding to the two possible signs marked with '$\pm$'.  We
mainly present results for unpolarized light, averaging over the two circular
polarizations. In addition to that, we decompose one selected spectrum
to two circular as well as to two axial polarizations.

\begin{figure*}[t]
\includegraphics*[width=2.0\columnwidth]{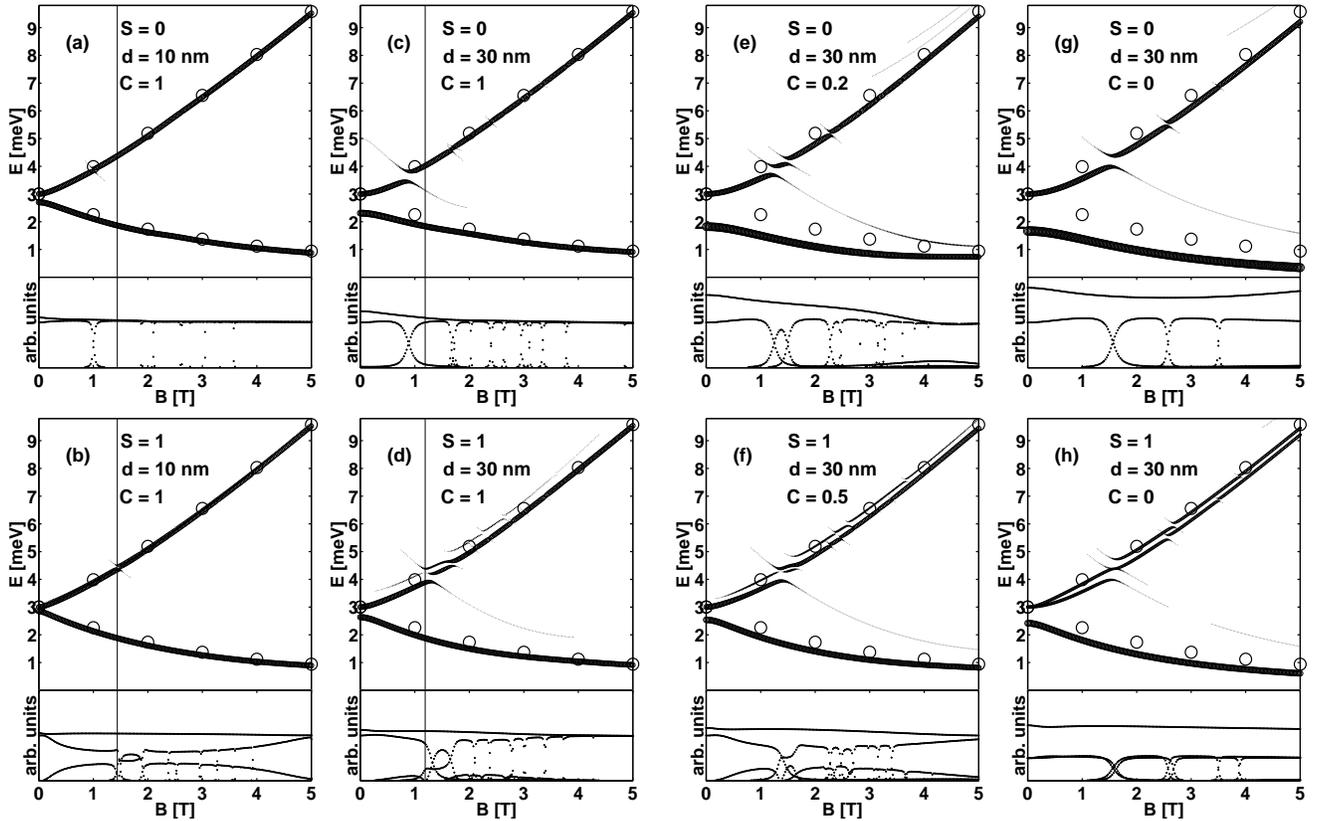}
\caption{\label{Fig:trans} 
FIR spectra of QDM with two different distances [(a) and (b):
$d=10$~nm, (e)-(h): $d=30$~nm] between QDs and full [(a)-(d)] and
reduced [(e)-(h)] Coulomb strengths $C$. Upper subfigures correspond
to $S=0$ spectra and lower $S=1$ one.  Vertical lines indicates the
transition points from spin-singlet ($S=0$) to spin-triplet ($S=1$)
state.  Spectra show the energy of absorbed light as a function of
magnetic field, and the width of lines indicate the transition
probabilities, also plotted separately below each spectrum. Circles
show the Kohn modes of an isolated parabolic QD.}
\end{figure*}

The calculated FIR spectra are shown in Fig.~\ref{Fig:trans} for
several different system parameters.  We plot the transition energy as
a function of the magnetic field using a line whose width is
proportional to the transition probability. We have only included the
transitions which have a probability of more than one percent of the
maximum value. To make it easier to compare the relative strengths of
the different resonances, we have plotted the probabilities also
separately below each spectrum.  In addition to that, we show the
energies of the two Kohn modes of a parabolic QD using open
circles. The upper row of figures present spectra for the total spin
$S=0$ and the lower one for $S=1$. The spectra for both spin types are
shown for two inter-dot distances $d$ and three Coulomb strengths $C$.
The ground state of our system changes from $S=0$ to $S=1$ when the
magnetic field is increased (see Ref.~[\onlinecite{Ari}] for more
details), and this transition is shown by a vertical line in the four
spectra of the fully interacting cases $C=1$. One should note that the
transition probabilities between different spins are zero, and thus
$S$ is conserved in the allowed transitions.

As a general feature of the calculated spectra shown in
Fig.~\ref{Fig:trans} one can see that each spectrum has as a major
component two branches, where the higher one ($\omega_+$) has a
positive dispersion and the lower ($\omega_-$) a negative one. For the
parabolic QD, these branches are the two Kohn modes (circles in the
figures). Now the lowered symmetry of the QD and the correlation
effects show up in the FIR spectra as a deviation from the Kohn modes.
These deviations are the splitting of $\omega_+$ and $\omega_-$ at
$B=0$, the anticrossings in $\omega_+$, and an additional mode
$\omega_{+2}$ above $\omega_+$ in the $S=1$ spectra. Next we will
analyze these deviations in detail.

Figs.~\ref{Fig:trans}~(a) and (b) show the ($C=1$) spectra for the
distance $d=10$~nm which corresponds to closely coupled QDs. One can
see that there are only minor deviations from the Kohn modes.  For
$S=0$, one can see a small $B=0$ splitting of the two modes. There are
also anticrossings in the upper modes, but these are hardly visible as
the energy gap is much smaller than the line width. These
anticrossings are, however, visible as oscillations in the transition
probabilities shown below the spectra. These probabilities also show
that the upper mode of the $S=1$ spectra actually consists of two
levels. These levels are energetically nearly degenerate, but the
probabilities vary as can be seen in lower part of
Fig.~\ref{Fig:trans}~(b). The sum of the two upper modes adds up to a
probability which is nearly constant and close to the probability of
the lower mode.

\begin{figure}[hbt]
\includegraphics[width=1.0\columnwidth]{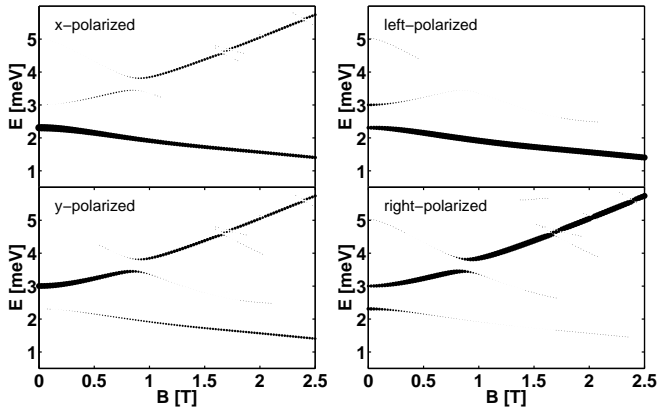}
\caption{\label{Fig:pol} Polarization dependence of FIR absorption for
a QDM with $d=30$ nm, $S=0$, and $C=1$. The left panel corresponds to
linearly polarized radiation and the right panel to two circular
polarizations.  At $B=0$ the modes are linearly polarized and for
large $B$ they approach circular polarizations.}
\end{figure}

If we increase the distance $d$ between QDs, the confinement potential
becomes less parabolic. In the extreme limit, however, the two QDs
decouple and one is left with two electrons localized in separated
dots. Figs.~\ref{Fig:trans}~(c) and (d) show the spectra for $d=30$~nm
which is in the interesting range of coupling between the two QDs. One
can see that the features discussed for $d=10$~nm are again present,
but at this time much clearer in the spectra. The $B=0$ splitting is
rather large for $S=0$, and there is also a small splitting in the
$S=1$ case.  There are many more anticrossings than in the $d=10$~nm
case, and those for weak $B$ now have a clear energy gap. The two
nearly degenerate modes of the $S=1$ case are now split in energy, and
the higher one ($\omega_{+2}$) has a smaller transition probability
than $\omega_+$. The comparison of the spectra of
Figs.~\ref{Fig:trans}~(a) to (c) and (b) to (d) shows that the first
anticrossing point is rather insensitive to the change of $d$ even if
the gap in energy opens up clearly. A further comparison shows that at
$B=0$, the $\omega_+$ mode stays at the energy of 3 meV, but
$\omega_-$ is lowered. The splitting of the modes results from the
fact that the symmetry in $x$- and $y$-directions is broken in the
confinement potential for a non-zero $d$. For our Hamiltonian, the
confinement along $y$-direction is kept parabolic even for $d>0$, and
the strength in this direction is 3 meV. For the $x$-direction, the
potential is non-parabolic for $d>0$, and the effective strength of an
approximating parabolic confinement is smaller than 3 meV for any
finite $d$.  In this way, the $B=0$ modes correspond to excitations to
either the short or long axis of the system. This fact can be further
demonstrated by considering the spectra of $x$- and $y$-polarized
radiation, presented in Fig.~\ref{Fig:pol} (a).
One can see that at $B=0$ the transitions are given purely by either
of the two linear polarizations. For a non-zero $B$, the two linear
polarizations mix due to a term $\mathbf{A} \cdot \nabla$ in the
Hamiltonian. Measurements using linearly polarized radiation lead to
similar conclusions, see Refs.~[\onlinecite{Dahl,Dahl2,Hochgrafe}].
For a strong magnetic field, the magnetic field length
($l_B=\sqrt{\hbar / eB}$) becomes smaller than the confinement length
($l_0=\sqrt{\hbar/ \omega_0 m^*}$) at $B \approx 1.7$ T.
Thus at greater $B$, the harmonic Landau potential dominates over the
QDM potential and the electrons start to localize in the minima of the
confinement potential~\cite{Ari}.  Therefore, the non-parabolicity is
diminished. The same effect can also be seen in the absorption of
circularly polarized radiation, shown in Fig.~\ref{Fig:pol}.

To separate the effects of the non-parabolic confinement potential
from those of interactions on the FIR spectra, we have performed
calculations with reduced Coulomb interaction strengths $0\le C <1$.
We present results for the $d=30$~nm case as it shows highly
non-trivial spectra. Figs.~\ref{Fig:trans} (g) and (h) show a
non-interacting spectra $C=0$, and Fig.~\ref{Fig:trans} (e) shows a
$S=0$ spectrum for $C=0.2$ and Fig.~\ref{Fig:trans} (f) the $S=1$ one
for $C=0.5$. Different intermediate values of $C$ are chosen for
different $S$ as the ones shown are the most representative ones.
Comparison of the $C<1$ spectra to those of Figs.~\ref{Fig:trans} (c)
and (d) with $C=1$ shows a very surprising feature: the spectra for
the reduced interaction strengths differ much more from the Kohn modes
than the fully interacting cases.
The splitting of the modes at $B=0$ grows as the interactions are
reduced, the anticrossing gaps are in general clearer, although a
direct comparison is not very easy as there are many shifts in the
anticrossing positions. In addition, the $\omega_{+2}$ mode of $S=1$
becomes more distinct in the spectra as $C$ is reduced.

We start the analysis of the $C<1$ spectra from the zero field
splitting of the two modes.
The potential along the $y$-axis is parabolic, and $\omega_+$  at
$B=0$ is not affected by the interactions. On the other hand, the
excitation energy along the non-parabolic axis is clearly influenced
by scaling $C$. The Coulomb repulsion effectively steepens the
confinement which leads to an increase in the excitation energy.

In experiments for elliptic QD lattices, the zero field resonance of
$\omega_{-} $ was found to exhibit an unexplained and clear dependence
on the number of electrons in the QD~\cite{Hochgrafe}. The lowest
energy was observed with the smallest number of electrons.
As the non-interacting $S=0$ spectrum coincides with the spectrum of a
one-electron QDM, one can study the particle number dependence of
$\omega_{-}$ for the smallest particle numbers.  Interestingly, we
find a very similar dependence in our data by comparing the
corresponding spectra in Figs. \ref{Fig:trans} (g) and (c).
The reduction of the $B=0$ excitation in Ref.~\cite{Hochgrafe} occurs
only in the long-axis direction. This implies a non-parabolic
potential in that direction, and on the other hand, a parabolic one in
the short-axis direction.

Another interesting question is how the anticrossings change when the
interactions are scaled down with $C<1$. 
For all interaction strengths, the energy gap at the anticrossing
point gets smaller as one moves to stronger $B$. 
This can be seen to result from the localizing effect of the magnetic
field: as the electrons localize around the potential minima, they
feel a more parabolic external confinement.
Now when the interactions are made weaker, the positions of the
anticrossings move, in general, to higher $B$. 
However, the energy gaps associated to fully and non-interacting cases
are nearly the same; the change is typically less than 10\%.
Thus the reduced interaction is compensated by the
stronger localizing effect of $B$ at the new anticrossing position of
higher $B$.
This can be understood by noting that interactions enhance
localization: the higher kinetic energy of localization is compensated
by the reduced interaction energy. In the extreme limit, a Wigner
molecule of electrons is formed~\cite{AriCL}.
It is now tempting to interpret this balance between the two
localizing effects in the way that the interactions affect the
anticrossings only indirectly via the effective potential.
Furthermore, as one cannot evidently identify any clear detail of the
spectra to result from the electron-electron interactions, the
possibility of FIR spectroscopy to reveal correlation effects is
somewhat questionable.  Its ability, however, to probe for the
effective confinement potential of electrons is clear. This is
especially true if various polarizations of radiation are used.

To summarize, we have presented FIR spectra of a lateral two-electron
quantum-dot molecule and studied the effects of confinement potential
and correlations on the spectra.
We have shown that lowering the symmetry of the confinement potential
induces changes in the spectra. These changes are deviations from the
two Kohn modes, and include the splitting of zero-field resonances,
anticrossings in $\omega_+$, and an additional mode above $\omega_+$
in the spin-triplet spectrum.
We have studied the effect of the electron-electron interactions on
these deviations, and surprisingly, the interactions drive spectra
towards those of a high-symmetry QD. All the deviations from Kohn
modes listed above are reduced as the interactions are turned on.
This leads to an interesting conclusion: the deviations from the Kohn
modes result merely from the non-parabolic external confinement
potential.  The interactions change the effective potential of the
electrons, but as our results show, the resulting potential is more
parabolic than the bare external confinement.  In this way, the
resulting FIR spectra are closer to Kohn modes than the
non-interacting ones.  In addition, our results show that probing
interaction effects by FIR spectroscopy is not straightforward, as
clear signatures of interactions are not present even in the
two-electron case where the interaction effects should be strongest.
On the other hand, FIR spectroscopy is able to reveal the
non-parabolicity of the external confinement potential very
efficiently, particularly if polarized radiation is used.

\begin{acknowledgments}

We acknowledge support by the Academy of Finland's Centers of
Excellence Program (2000-2005).
  
\end{acknowledgments}

\end{document}